\documentstyle[epsfig]{aipproc}
\begin{document}
\title{Looking into the Future of Spin and QCD}

\author{Jacques Soffer}
\address{Centre de Physique Th\'eorique\\
CNRS Luminy Case 907\\
13288 Marseille Cedex 09 France}

\maketitle

\begin{abstract}
Several stimulating open questions in high energy spin physics will be
described together with the striking progress recently achieved in this
field. In view of the new experimental facilities and the new tools, soon
available, I will try to anticipate what we will learn next. The prospects for
the future are excellent and clearly, some exciting times are ahead of us.
\end{abstract}

\section*{Introduction}
Although this is the concluding lecture of SPIN2002, I was asked NOT to 
give a summary talk, but rather to " look into the future ", in other words
to foresee, for example, what will be the highlights of SPIN2010! Soon I
realized it is not an easy task, but it was too late because I had already
accepted it. Let me first focus on one of the key words and quote twice a great man,
A. Einstein who said:

- {\it I never think about the FUTURE, it comes too soon}

- {\it To make predictions is very difficult, mainly when it concerns the
FUTURE}

In preparing this lecture I was thinking "I need to learn how to read off
the crystal ball" and I came across the existence of the {\it Institute for
the Future, 2744 Sand Hill Rd. Menlo Park, CA 94025}, which didn't help.\\
Turning now to physics questions, I want to stress the fact that {\it spin}
is a powerful and elegant tool which will continue, as in the past, 
to play a crucial role in physics. One of the main goals of high energy
particle physics is to understand the fundamental structure
of hadrons. In particular concerning the nucleon, which is composed of quarks,
antiquarks and gluons, it is essential to determine the role of each parton
in making the nucleon spin. It is widely accepted that hadron dynamics is
accurately described by perturbative QCD, which has been tested so far to
a certain level of accuracy. However it is urgent to improve it and to
confirm its validity, by means of new spin experiments which will test, for the 
first time, the spin sector of QCD.

The future of high energy physics lies mainly in discovering physics beyond
the Standard Model and this is the essential motivation for building a multi-TeV
proton collider, like LHC, or new $e^+e^-$ linear colliders. However, spin can
be also very relevant for searching new physics, which might exist at the
TeV energy scale. One way is to detect indirect effects at low energy, by 
measuring spin-dependent observables with a very high precision and we will present 
a few specific cases. Another way is to use several hundreds GeV polarized proton beams,
like at RHIC, to carry out a general spin physics programme \cite{BSSV} and 
to uncover some direct effects in $pp$ collisions, as we will see. One should not forget 
the prospects we also have with polarized $e^+e^-$ linear colliders \cite{GMP}.

Before I go on, I would like to apologize for leaving out so many interesting topics
because of my ignorance or by lack of time.

\section*{ Windows for physics beyond the Standard Model}
\begin{center}
{\bf Neutron Electric Dipole Moment}
\end{center}
The existence of a neutron electric dipole moment (EDM) $d_n$ requieres parity P and time
reversal T violation and the latter one implies CP violation, due to the
CPT theorem which has very strong physics grounds. So a natural way to
improve our understanding of the origin of CP violation is to search for
non-zero $d_n$. In the Standard Model, predictions from the 
Cabibbo-Kobayashi-Maskawa mixing matrix give the upper
limit $d_n < 10^{-31} e$cm, which is around six orders of magnitude from the 
present experimental limits $d_n < (6-10) \cdot 10^{-26} e$cm. However beyond
the Standard Model there are many sources of CP violation (Susy, Strong CP,
electroweak baryogenesis, etc...)
which lead to $d_n < 10^{-26} e$cm. A new experiment is planned at 
Los Alamos Nat.Lab. \cite{REM}, whose goal is to improve the sensitivity, in order to push
the limit to $10^{-27} e$cm by 2004 and to $10^{-28} e$cm, ten years from now.

\begin{center}
{\bf Parity Violating Asymmetries}
\end{center}
The most accurate determination of the electroweak mixing angle $sin^2\theta_w$, at the
energy scale of the $Z$ pole, comes from the SLC and LEP experiments. This fundamental
physical constant is running with $Q^2$, according to the radiative corrections, so its
value at lower $Q^2$ is well predicted by the Standard Model. This can be tested, for 
example, by measuring parity violation in M{\o}ller scattering, as planned by the E158 
experiment at SLAC \cite{DRR}. They are using a 50GeV polarized electron beam which is 
scattered off unpolarized atomic electrons and measure the asymmetry $A_{PV}=
(\sigma_R - \sigma_L)/(\sigma_R + \sigma_L)$. It is expected to be very small, of
the order of $10^{-7}$, since it goes like $(1/4 - sin^2\theta_w)$. The goal of E158
is to reach a precision $\delta sin^2\theta_w = \pm 0.0008$ by the end of 2003. With this
achievement, the sensitivity to new physics is up to $\Lambda = 15$TeV for compositeness, 
for a new boson $Z'$ of mass $M_{Z'}=1$TeV and also for lepton flavor violation, to some
extend. We look forward to the impact of the results of this new challenging experiment.

\begin{center}
{\bf The Anomalous Magnetic Moment of the Muon}
\end{center}
Another way to probe possible extensions of the Standard Model is to make a high
precision measurement of the anomalous magnetic moment of the muon. It is characterized
by the value of $a_{\mu}=(g-2)/2$, which vanishes for the case of a pointlike
elementary Dirac particle since $g=2$. For the muon the theoretical value of $a_{\mu}$
in the Standard Model is obtained by adding up QED, hadronic and weak contributions
and the current most recent estimate is $a_{\mu}(SM) = 11~659~177(7)\cdot 10^{-10}$
, predicted with an accuracy of 0.6ppm. A twenty years old CERN experiment on positive
muon had obtained an accuracy of only 10ppm and this has motivated a new experiment
at BNL (E821), which has recently reported \cite{E821}$a_{\mu^+} = 11~659~204(7)(5)
\cdot 10^{-10}$, a value with an accuracy of 0.7ppm. This experiment plans to reduce
the uncertainty to about 0.35ppm and the data obtained so far for negative muon, is
also expected to be released soon. On the theoretical side one should try to improve
the understanding of QCD corrections and the use of better data (for example on $e^+e^-$
collisions) in order to reduce the still rather large uncertainty on hadronic corrections. We
note that $a_{\mu^+}^{exp}>a_{\mu^+}^{th}$, so may be Susy is round the corner!

\section*{Spin in elastic scattering}

\begin{center}
{\bf Spin Properties of the Pomeron}
\end{center}
In the framework of Regge phenomenology, two-body hadronic reactions at very high
energy $s$ and small momentum transfer $t$, are described by the Pomeron exchange.
This important trajectory, whose fundamental nature is not yet fully elucidated,
couples to hadrons and the forward nonflip amplitude $F^{\it P}_{\it nf}$ is directly
related to the total cross section, by the optical theorem. For simplicity, one
usually assumes that the flip coupling of the
Pomeron vanishes, so for the single flip amplitude, one has $F^{\it P}_{\it sf}=0$.
This is theoretically groundless and essentially based on $s$-channel helicity
conservation, in a perturbative treatment of the Pomeron. 
It should be checked, for example in the forward
direction, by measuring total cross sections in pure spin states and for $t\neq 0$
by measuring spin dependent observables, like the analyzing power (or left-right
asymmetry) $A_N$, in $pp$
elastic scattering. This is by all means an interesting physics programme for the
pp2pp experiment at RHIC. A good knowledge of the spin properties of the Pomeron has
also relevant implications for polarimetry at high energy, a crucial issue for
spin measurements at RHIC. One possible method is to measure $A_N$ in proton-proton
or proton-nucleus elastic scattering \cite{BK}, at very small momentum transfer
$10^{-3}\leq -t \leq 10^{-2}$, the so-called Coulomb-nuclear interference (CNI) region.
It originates from the interference between the hadronic non-flip amplitude $F^{\it had}_{\it nf}$
and the electromagnetic (Coulomb) spin-flip amplitude, which are out of phase.
The theoretical prediction depends strongly on the existence of a spin-flip
hadronic amplitude $F^{\it had}_{\it sf}$. This contribution is characterized
by the ratio $r_5 = (m_p/\sqrt{-t}) F^{\it had}_{\it sf}/ImF^{\it had}_{\it nf}$ and
a new BNL-AGS experiment (E950), which has measured $A_N$ in proton-carbon elastic
scattering in the CNI region at 21.7GeV/c, has obtained \cite{CNI} $Rer_5 = 0.088 \pm 0.058$
and $Imr_5 = -0.161 \pm 0.226$. Although it is tempting to associate this with a spin-flip
contribution of the Pomeron, one needs to wait for further results at higher RHIC energies with
better accuracy.

\begin{center}
{\bf Nucleon Electromagnetic Form Factors}
\end{center}

The electromagnetic form factors of the proton contain important informations on its internal
structure and they can be measured in electron-proton elastic scattering. However for the
region of large four-momemtum transfer squared, say $Q^2 > 1 \mbox GeV^2$, 
from the unpolarized cross section, it is not possible
to separate the electric form factor $G_{Ep}(Q^2)$ from the magnetic form factor $G_{Mp}(Q^2)$,
which gives the dominant contribution. $G_{Mp}(Q^2)$ is known to behave approximately
as a dipole $[1 + Q^2/0.71]^{-2}$ and the same behavior was assumed for $G_{Ep}(Q^2)$, such
that $\mu_p G_{Ep}(Q^2)/G_{Mp}(Q^2) = 1$, where $\mu_p$ is the proton magnetic 
moment. In fact this ratio of form factors can be measured using the recoil polarization 
technique, because it is proportional to the ratio of the transverse to longitudinal components
of the recoil proton in the elastic reaction $ \overrightarrow e p \rightarrow
e \overrightarrow p$. According to a recent JLab experiment \cite{FF}, one finds that
the above naive assumption is not true and one has, instead, $\mu_p G_{Ep}(Q^2)/G_{Mp}(Q^2) = 1 - 
0.13~(Q^2 - 0.04)$, up to $Q^2 = 5.6 \mbox GeV^2$. If one extrapolates this linear trend
to higher $Q^2$, the electric form factor should exhibit a zero near $Q^2 = 8 \mbox GeV^2$ and
this measurement will be done in the near future.
The theoretical implications of this result
in the framework of various models were discussed by J.J. Kelly \cite{JJK}, who also 
presented some new results for the neutron electromagnetic form factors.

\section*{ Is the riddle of the proton spin structure solved ?}

This basic question is with us for nearly 15 years and although hudge
progress have been made, both on experimental and theoretical sides, we don't
have yet the final answer. Let us recall the fundamental proton spin sum rule
\begin{equation} \label{1}
1/2 = 1/2 \Delta \Sigma + \Delta g + L_q + L_g
\end{equation}
where $\Delta \Sigma = \sum_{i}(\Delta q_i + \Delta \bar q_i)$ is the fraction
of the proton spin carried by quarks and antiquarks. The summation runs over 
all the different flavors, $u,d,s,...$, and all these quantities are first
moments of the corresponding $x$-dependent distributions, $\Delta g$ is the spin
carried by the gluon and $L_{q,g}$ are the quark and gluon orbital angular
momentum contributions. In the naive parton model $\Delta g = 0$ and one ignores
$L_{q,g}$, so in order to fulfill the above sum rule one has $\Delta \Sigma = 1$.
However in 1988 the EMC data \cite{EMC} led to a small $\Delta \Sigma$, a big surprise
which was interpreted as the fact that the sea quarks carry a fraction of
the proton spin, negative and large, so to make effectively $\Delta \Sigma \sim 0$.
This inclusive polarized Deep Inelastic Scattering (DIS) data has been greatly improved
since then, not only for the proton but also for the neutron, but the earlier trend
remains, since we have now $\Delta \Sigma = 0.23 \pm 0.04 \pm 0.06$ at $Q^2 = 5 \mbox GeV^2$ 
\cite{TA}. One should keep in mind that in this type of experiment one measures the
asymmetry $A_1(x,Q^2)$, from which one extracts the spin-dependent structure function
$g_1(x,Q^2)= {1/2}\sum_{i}[\Delta q_i(x,Q^2) + \Delta \bar q_i(x,Q^2)]$. Clearly this does
not allow the flavor separation, or to disentangle quarks from antiquarks, and some
attempts to achieve it were done by measuring semi-inclusive polarized DIS \cite{MA}.
The HERMES Collaboration has produced some interesting new results showing that, at  
least in the limited kinematic region they have investigated, the sea quarks contribution 
is not large and negative. Within a rather low accuracy, there is perhaps an indication 
for $\Delta s(x) > 0$ and the data are consistent with flavor symmetry, 
{\it i.e.} $\Delta \bar u(x) \sim \Delta \bar d(x)$. This can be better checked in the
future at RHIC, by measuring parity-violating helicity asymmetries $A_L^{PV}$ in $W^{\pm}, Z$ 
production. This idea was first proposed nearly 10 years ago \cite{BS} and according
to the Drell-Yan production mechanism, one gets simple expressions for these asymmetries.
The results of recent calculations \cite{BBS} for $A_L^{PV}(W^{\pm})$ are shown in Fig.~1.

\begin{figure}[ht]
\begin{center}
\leavevmode {\epsfxsize=6.5cm \epsffile{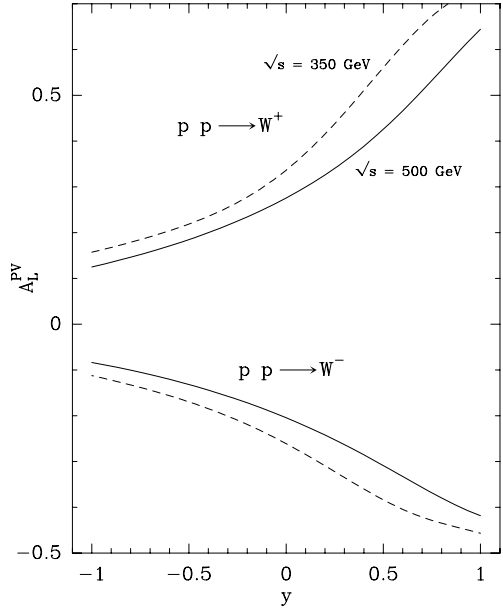}}
\end{center}
\caption[*]{\baselineskip 1pt
The parity violating asymmetry $A_L^{PV}$
for $ p p \rightarrow W^{\pm}$ production versus the $W$ rapidity
at $\sqrt{s} = 350 \mbox{GeV}$ (dashed curve)
and $\sqrt{s} = 500 \mbox{GeV}$ (solid curve)~ (Taken from Ref.~\cite{BBS}).}
\label{Fig1}
\end{figure}
Their general trend can be easily understood and, for example at $\sqrt s = 500\mbox{GeV}$  
near $y=+1$, $A_{L}^{PV}(W^+) \sim \Delta u /u$ and
$A_{L}^{PV}(W^-) \sim \Delta d /d$, evaluated at $x=0.435$. 
Similarly for near $y=-1$, $A_{L}^{PV}(W^+) \sim -\Delta \bar d /\bar d$ and
$A_{L}^{PV}(W^-) \sim -\Delta \bar u /\bar u$, evaluated at $x=0.059$. 
Given the expected rates
for $W^{\pm}$ production at RHIC-BNL and the high degree of the 
proton beam polarization \cite{BSSV}, it will
be possible to check these predictions to a high accuracy.

Although, since 1988 a major effort has been made to explore the small and medium
$x$-regions, for the purpose of checking the fundamental Bjorken sum rule \cite{Bj}, it is
also important to improve the existing high $x$ data on the spin dependent 
structure functions. The preliminary results from a JLab experiment (E99-117) were 
presented \cite{XZ} and they clearly show that $A_1^n(x)$ changes sign for $x\sim 0.45$
and becomes positive and large for $x\sim 0.6$. This important feature will have to be 
confirmed in the future, up to $x\sim 0.75$ by JLab upgrade to 12GeV.

Finally, we still have a very poor knowledge on the contribution of the gluon to the
proton spin, which is very badly constrained by the QCD fits to the available polarized
DIS data, whose $Q^2$ coverage is rather limited. There are various direct ways to
get access to $\Delta g(x)$ and, for example the COMPASS experiment at CERN will do it
by means of charm particles production, through the photon-gluon process 
$\gamma g \rightarrow c \bar c$, with $c$ ($\bar c$) hadronizing into $D^0$ ($\bar D^0$).
The final uncertainty on the measurement of $\Delta g(x)/g(x)$ from open charm is estimated
to be $\delta (\Delta g(x)/g(x)) =0.11$, in the $x$-range $0.02 < x < 0.4$ at $Q^2=10 \mbox GeV^2$.
The HERMES experiment at DESY is also expected to collect some data with a lower accuracy, in a smaller
$x$-range and at lower $Q^2$.
 
There are also several very promising methods for extracting $\Delta g(x)$ in polarized $pp$ 
collisions at RHIC, by measuring the double helicity asymmetry $A_{LL}$. Some of the key 
processes which will be investigated, both at $\sqrt{s}=200 \mbox GeV$ and $500 \mbox GeV$, are
high-$p_T$ prompt photon production 
$pp\rightarrow \gamma + X$, jet production $pp\rightarrow jets + X$ and heavy flavor
production $pp\rightarrow c \bar {c} + X$, $pp\rightarrow b \bar {b} + X$, $pp\rightarrow J/\psi + X$.
Carefull studies about the expected accuracy have been completed recently \cite{MS} and
it was also proposed to get very soon, some relevant determination of $\Delta g(x)$ by looking at the
production of leading high-$p_T$ pions, which are copiously produced even with a moderate
luminosity. In all these reactions the subprocesses are either $g q \rightarrow \gamma q$,
$g q \rightarrow g q$, $g g \rightarrow g g$, etc.., for all of which, the double helicity asymmetry
$ a _{LL}^{ij}$ is positive. Therefore the sign of the predicted $A_{LL}$ is always positive
because there is a strong prejudice to anticipate $ \Delta g(x) > 0 $. However one can speculate
that Susy particles are not too heavy \cite{GMP}, so that they could be produced in the above
processes. In this case, one should remember that the subprocesses 
$g g \rightarrow \tilde g \tilde g$, $g q \rightarrow \tilde g \tilde q$, etc.., all have their
${\tilde a} _{LL}^{ij}$ are negative and large, -100\% or so, and this could reduce the effect,
leading to a misinterpretation of the experimental results.

Clearly when a precise determination of $\Delta \Sigma$ and $\Delta g$ will be obtained, 
by using the spin sum rule Eq. (\ref{1}), it will be possible to estimate the orbital angular momentum
contributions, which are unknown so far. According to recent theoretical developments, they are
related also to generalized (off-forward) parton distributions (OFPD), which might be measured, for
example, in deeply virtual Compton scattering (DVCS), where a real photon is produced and in 
diffraction meson production \cite{MV}.

\section*{Transverse spin physics}
\begin{figure}[ht]
\begin{center}
\leavevmode {\epsfxsize=7cm \epsffile{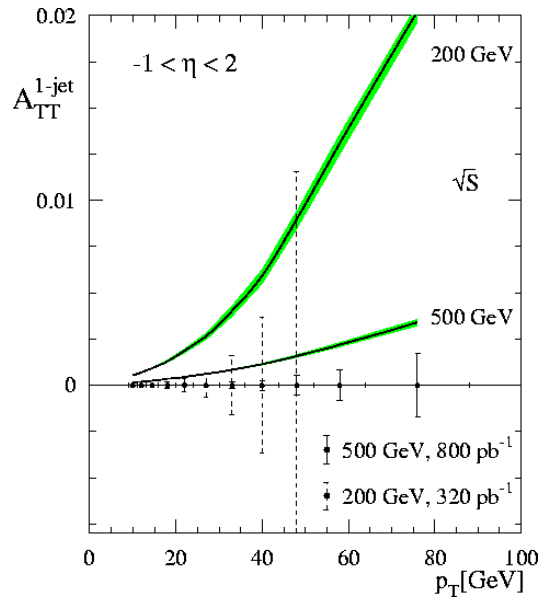}}
\end{center}
\caption[*]{\baselineskip 1pt
Upper bounds for $A_{TT}$ for single-inclusive jet production
at RHIC versus $p_T$ for $\sqrt{s}=200 \mbox GeV$ and $500 \mbox GeV$ (Taken from 
Ref. \cite{SSV}).}
\label{Fig2}
\end{figure}

\begin{center}
{\bf Quark Transversity in a Nucleon}
\end{center}
In addition to the unpolarized and polarized quark distributions $q_i(x,Q^2)$ and
$\Delta q_i(x,Q^2)$, discussed so far, there exists a third quark distribution, also
at leading-twist order, called transversity and denoted $h_1^{q_i}(x,Q^2)$ or 
$\delta q_i(x,Q^2)$. It is not accessible in DIS because it is chiral-odd and
there is no corresponding transversity for gluons, due to helicity conservation \cite{PR}. The
quark transversity can be measured, in conjonction with a chiral-odd fragmentation
function, in the single-spin azimuthal asymmetry for inclusive pion production
$ e p^{\uparrow} \rightarrow e \pi X $, with a transversely polarized target. There is a first
indication of the effect in the HERMES data \cite{AA} on $A_{UL}(\phi)$, with a longitudinally
polarized target. Another possibility is to measure the double-transverse asymmetry $A_{TT}$
in Drell-Yan dimuon production in $pp$ collisions at RHIC. To this process $p^{\uparrow}p^{\uparrow} 
\rightarrow \mu^+ \mu^- X$, corresponds an asymmetry which involves the product 
$h_1^{q_i}(x,Q^2) \cdot h_1^{\bar q_i}(x,Q^2)$. The calculation of $A_{TT}$ has been done at 
$\sqrt{s}=200 \mbox GeV$ in next-to-leading order \cite {MSSV} and using the positivity bound on
transversity distributions \cite{JS}, but it leads to an effect of a few precents, which will
not be easy to detect by lack of statistics.
Finally, it is important to notice the small size of double-transverse 
asymmetries in other reactions like jet production or prompt photon production. We anticipate
a strong dilution of $A_{TT}$ in the gluon induced processes and such an example is displayed 
in Fig. \ref{Fig2}. So, in future measurements, we should always observe $|A_{TT}| << |A_{LL}|$,
 a definite test of QCD.

\newpage

\begin{figure}[ht]
\begin{center}
\leavevmode {\epsfxsize=13cm \epsffile{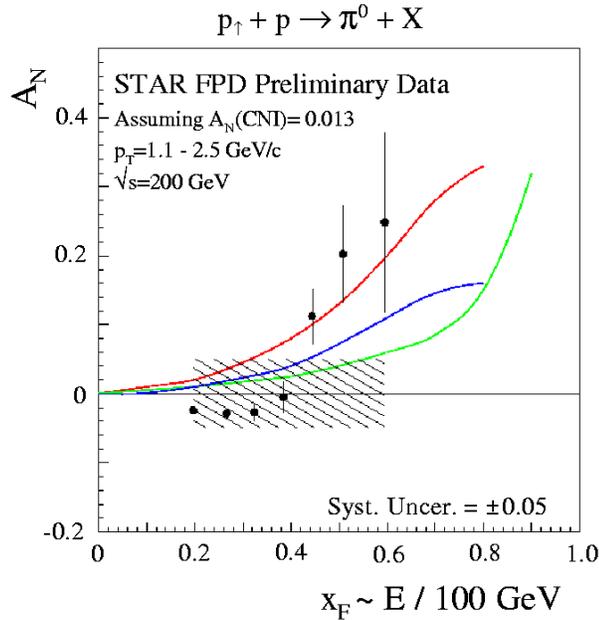}}
\end{center}
\caption[*]{\baselineskip 1pt
Preliminary data from STAR at RHIC and theory predictions at $p_T= 1.5 \mbox GeV/c$
(All three are private communications). Upper curve, Collins effect \cite{AM99}-Middle
 curve, Sivers effect \cite{AM98}-Lower curve, Twist-3 effect \cite{QS}.}
\label{Fig3}
\end{figure}

\begin{center}
{\bf Single Transverse Spin Asymmetries}
\end{center}
The study of single transverse spin asymmetries, usually denoted $A_N$ as above, in 
one-particle inclusive hadron collisions $a + b \rightarrow c + X $ (with either $a$, $b$ or $c$,
transversely polarized) is strongly motivated by large effects observed in
a vast collection of intriguing data, for 
example, in hyperon production \cite{LP} or in single pion production \cite{DLA}. 
According to naive theoretical arguments, they were expected to be zero for several years. 
They are non-zero {\it only}
if there is an interference between a non-flip and a single-flip amplitudes, out of phase.
Several possible mechanisms have been proposed recently \cite{PR} and, in this respect,
 I would like to give a warning: make sure that the proposed mechanism to explain $A_N$ 
is also appropriate to describe, {\it  both in shape and magnitude}, the corresponding unpolarized 
cross section in the {\it same kinematic region}, because it contains the bulk of the
underlying dynamics. We show in Fig. \ref {Fig3} some preliminary results obtained at
$\sqrt{s}= 200 \mbox GeV$ at RHIC \cite{GR}, which confirm the trend of the FNAL E704 data \cite{DLA} at
$\sqrt{s}= 19.4 \mbox GeV$. We look forward to the future and to some results at higher $p_T$, which
will strongly test the different theoretical models, provide the authors dare to make predictions
in this new kinematic region, prior to the release of the data.

\section*{Hunting for new physics at RHIC}
Like we said earlier, the use of high energy polarized proton beams at RHIC is a new
tool to try to uncover new physics. As a specific example, let us consider the parity
violating asymmetry $A_L^{PV}$, in single-jet production. A Standard Model asymmetry
is generated from a QCD electroweak interference and this contribution, at $\sqrt{s}= 500 \mbox GeV$,
gives a positive asymmetry which rises, with the transverse momentum of the jet, up to 
 2-3 \% for $p_T \sim 150 \mbox GeV$, with a small uncertainty related to that of the
parton distributions. However, instead of this small effect, one could
observe sizable effects from contact interactions with a new energy scale $\Lambda > 1 \mbox TeV$
or a new vector boson $Z^{'}$ with a mass up to $1 \mbox TeV$ \cite{TV}. It was shown that RHIC
can be competitive with Tevatron run II, because high luminosity and polarization
give a larger sensitivity despite the lower energy. RHIC is also a unique place to probe 
the chiral structure of new interactions.

Another example is the possibility to detect a large CP-violation in lepton pair
production in polarized $pp$ collisions at RHIC. Let us consider the single-transverse
asymmetry $A_T$ in the reaction $p^{\uparrow} p \rightarrow {\it l}^{\pm} \nu X$, where
${\it l}^{\pm}$ is a charged lepton. It was shown that some mechanisms beyond the Standard
Model can generate a non-zero asymmetry, which was estimated to be rather large in
phenomenological tensor interactions \cite{KSS}. This interesting speculation should
be checked in future experiments at RHIC.\\

To complete the exciting talk we heard this morning on {\it Applications of Spin in Other Fields} 
\cite{GC}, let me mention one recent piece of work, where a {\it Soft Spin Model} is used to study the
fluctuations and market friction in financial trading \cite{BR}, a subject of great interest! 
There are also some topics which may belong to "Science Fiction and the Future of Spin", 
like Quantum information, Quantum teleportation or Quantum computation. No doubt that a successful 
quantum computer would be exponentially faster than a classical computer.

Finally, let me refer to a nice lecture "History of Spin and Statistics" by A. Martin \cite{AM} 
who stressed the fact that the stability of the world is due to the electrons having 
a spin-1/2. In agreement with a statement made in the opening talk of SPIN2002 \cite{XJ},
 I also claim:\\

{\it A world without spin would collapse: so keep it UP (or DOWN as you wish)!}
 
\section*{Acknowledgments}
I would like to thank the organizers, in particular Yousef Makdisi, for the opportunity
to deliver this closing lecture at the very successful conference "SPIN2002". I also thank
many participants of this meeting for helpful discussions.

\end{document}